%% file: main.tex
\documentclass[runningheads]{llncs}
\usepackage[T1]{fontenc}
\usepackage{graphicx}
\usepackage{amsmath,amsfonts}
\usepackage{algorithmic}
\usepackage{algorithm}
\usepackage{array}
\usepackage{subfigure}
\usepackage{booktabs}
\usepackage{longtable}
\usepackage{hyperref}
\usepackage{url}
\usepackage{tabularx}
\usepackage{multirow}

\begin{document}
\sloppy
\title{Llama-Polya: Instruction Tuning for Large Language Model based on Polya's Problem-solving}
\author{
Unggi Lee$^{1,\dagger}$ \and
Yeil Jeong$^{2}$ \and
Chohui Lee$^{3}$ \and
Gyuri Byun$^{4}$ \and
Yunseo Lee$^{5}$ \and\\
Minji Kang$^{6}$ \and
Minji Jeon$^{7,\dagger}$
}

\authorrunning{U. Lee et al.}

\institute{
$^1$Korea University Sejong Campus, $^2$Indiana University Bloomington\\
$^3$Ewha Womans University, $^4$Seoul National University,\\
$^5$University of Wisconsin-Madison, $^6$Daegu Gyeongdong Elementary School\\
$^7$University of Nebraska-Lincoln\\[6pt]
\email{codingchild@korea.ac.kr, yeilj@iu.edu}\\
\email{etdev97@ewha.ac.kr, gl3013@snu.ac.kr, ylee899@wisc.edu}\\
\email{buruburu725@gmail.com, mjeon3@unl.edu}\\[2pt]
$^\dagger$Corresponding authors
}

\maketitle

\input{0_abstract}
\input{1_introduction}
\input{2_literature_review}
\input{3_method}

\input{4_result}

\input{5_discussion}
\input{6_conclusion}

\input{7_appendix_survey}

\end{document}

%% file: 0_abstract.tex
\begin{abstract}
This paper introduces Llama-Polya, an instruction-tuned large language model that integrates Polya’s four-step problem-solving framework into its dialogue structure to support mathematical reasoning. Mathematical problem-solving is central to students’ success in mathematics education, yet many learners struggle to plan, justify, and verify their solutions. Although large language models (LLMs) show promise as intelligent tutors, they often lack structured pedagogical alignment grounded in established learning theories.

To address this gap, we operationalize Polya’s problem-solving framework within an instruction-tuned LLM to promote metacognitive engagement and examine the effects of pedagogy-aligned fine-tuning compared to domain-only and general-purpose instruction tuning. Built on the Llama-3.1-8B architecture, Llama-Polya was fine-tuned on synthetic math problem-solving data derived from GSM8K, structured according to Polya’s four stages. We developed and evaluated multiple variants—general-purpose instruct, math-domain metamath, pedagogy-aligned polya-v2, and sequential metamath+polya-v2—using both quantitative accuracy metrics and qualitative pedagogical assessments.

Results indicate that models tuned with Polya’s framework and domain-specific data produced more balanced reasoning-stage distributions and fewer premature answers. Expert evaluators also observed improved pedagogical coherence and metacognitive prompting, although limitations in personalization and mathematical rigor remained. These findings suggest that pedagogy-grounded instruction tuning can enhance educational alignment and reasoning transparency in LLM-based tutoring systems.
\end{abstract}

%% file: 1_introduction.tex
\section{Introduction}

Problem-solving is essential in math education and a driver of students' cognitive development and achievement \cite{henderson1953problem,goldin1997chapter}. To cultivate this competency, Polya's four-step method---understanding the problem, devising a plan, carrying out the plan, and looking back---has long informed instructional design and assessment for math problem-solving \cite{polya2004solve}. However, many learners still struggle to grasp the context of problems, connect mathematical concepts, and independently build a strategy to find the answer \cite{yapatang2022development,duval2006cognitive,lee2016appropriate}.

Large language models (LLMs) offer a path to address these challenges in math education by enabling interactive, context-aware tutoring that affords personalized learning \cite{holmes2023artificial,luckin2017towards}. However, many models are tuned for broad tasks rather than pedagogy-aligned guidance responsive to learners' needs \cite{wardat2023chatgpt,baidoo2023education}. Instruction tuning can narrow this gap by aligning model behavior with specific instructional goals \cite{ouyang2022training,chung2024scaling,kasneci2023chatgpt}, but applications to math problem-solving in educational contexts remain underexplored.

We introduce Llama-Polya, an instruction-tuned LLM that operationalizes Polya's method and provides personalized scaffolding through multi-turn dialogue. Rather than supplying solutions, the model elicits understanding, plans, justifications, and reflective checks, capturing students' difficulties, providing step-by-step guidance, and hence supporting their mathematical understanding and critical thinking process. We designed a prompt‑to‑dialogue data pipeline encoding stage‑specific goals and utterance guidelines, fully fine‑tuned Llama‑3.1‑8B on synthetic tutoring dialogues derived from GSM8K, and evaluated variants with stage‑wise annotations and expert judgment across arithmetic, measurement, and geometry.

With the introduction of Llama-Polya, this study aims to contribute to the growing body of knowledge on AI applications in education and explore the potential of instruction-tuned LLMs for supporting mathematical problem-solving. By grounding LLMs in an established pedagogical framework, Llama-Polya offers a more adaptive and pedagogically meaningful approach that supports both educators and learners \cite{kumar2023math,matzakos2023learning}.

%% file: 2_literature_review.tex
\section{Literature Review}

\subsection{Mathematics Problem-Solving}

In education, mathematical problem-solving engages learners in reasoning beyond rote application of procedures by interpreting contexts, planning, executing multi-step work, and evaluating solutions \cite{henderson1953problem,goldin1997chapter}. Curriculum perspectives emphasize its centrality for developing communication, creativity, and reasoning \cite{national1980agenda,tambunan2019effectiveness}, positioning problem-solving as a core outcome of math education. Persistent learner challenges poses threats to developing problem-solving abilities, particularly in concept application, abstract reasoning, and self-monitoring \cite{yapatang2022development}, often lead students to revert to formulaic recall rather than coordinated reasoning \cite{duval2006cognitive,lee2016appropriate}.  Polya's Four-Step Process of problem-solving offers systematic support for designing math instruction and assessment. In Polya's method, (1) learners first understand the problem by clarifying givens, unknowns, and constraints; they then devise a plan by selecting and justifying a strategy and outlining necessary calculations or constructions; next, they carry out the plan, executing steps while monitoring progress; and finally they look back, verifying the result, relating it to the original question, and reflecting on both the solution path and their understanding and problem-solving process. 

Building on this foundation, subsequent scholarship elaborated the protocols and metacognitive processes that sustain Polya's work. Schoenfeld \cite{schoenfeld1981episodes} described problem-solving episodes---reading, analysis, exploration, planning and implementation, and verification---interwoven with executive decisions about goals and resources. Garofalo and Lester \cite{garofalo1985metacognition} identified cognitive-metacognitive categories---orientation, organization, execution, and verification---that map onto and extend Polya's phases for diagnostic and instructional purposes. These frameworks render the four steps observable as instructional activities and clarify where and how teachers can scaffold.  Instructional approaches incorporating prompts based on Polya's stages have demonstrated effectiveness in enhancing problem-solving abilities across achievement levels, by providing structured support \cite{lee2016appropriate,ukobizaba2021assessment}. 

Yet, these gains do not automatically carry over to digital implementations as the translated computer-assisted systems, with its over-reliance to programmed solutions and extensive resources, can become over-scaffolding, deliver generic or inaccurate feedback, misaligned with learners' needs, or channel exploration into predetermined schemas, narrowing the scope of self-directed learning \cite{chang2006computer}. 
This points to a concrete design requirement on providing phase-aware guidance that is sensitive to where the learner is in the Polya's four step and precise about the next move forward.

\subsection{Artificial Intelligence in Mathematics Education}

Digital technologies have supported learning math through modeling, visualization, and collaboration \cite{bray2017technology,drijvers2010introduction,olive2010mathematical}, with teachers leveraging tools to provide more timely and meaningful feedback \cite{haddad2002technologies,geiger2012technology,kaput2020technology}. AI systems extend these capabilities by diagnosing performance \cite{hwang2021roles}, tracing knowledge over time e\cite{casalino2021deep}, and recommending tasks \cite{cunska2020effective}. Yet many tools fail to address students' learning histories and processes holistically, making adaptivity to evolving strategies and misconceptions difficult to achieve \cite{casalino2021deep,wardat2023chatgpt}. 

Several pedagogical lenses inform AI use in classrooms. Knowledge building frames learning as the progressive refinement of ideas, highlighting student explanation and questioning \cite{roscoe2007understanding}. Feedback design emphasizes actionable, timely information that supports regulation and improvement, not just judging accuracy \cite{dawson2023technology}. Distributed cognition conceptualizes thinking as spanning internal and external structures, opening a path toward student-AI collaboration where tools share cognitive load and mediate strategy use \cite{kim2022learning}. 
Within this landscape, large language models (LLMs) are being explored for content generation and conversational support in math education.

Empirically, LLMs have shown utility for instructional content generation, for example, drafting teaching plans \cite{hu2024teaching} or generating age-appropriate word problems \cite{christ2024mathwell}, which reduces design overhead for educators \cite{feng2024exploring,matzakos2023learning}. They can also act as conversational tutors, offering step-by-step support that can improve learner performance \cite{baidoo2023education,matzakos2023learning}, and lower perceived difficulty for certain tasks \cite{kumar2023math}. However, persistent concerns include response accuracy, alignment with pedagogical intentions, and personalization to diverse learners and contexts \cite{kumar2023math,wardat2023chatgpt}. These concerns point to the need for pedagogy-aligned tuning and evaluation methods that target process-level guidance and constructive feedback rather than only final answers.

\subsection{Instruction Tuning of Large Language Models}

Instruction tuning aligns pretrained LLMs to follow targeted behaviors and styles. Early work demonstrated that fine-tuning on curated instruction-response pairs and, in some cases, human feedback, substantially improves task following and user-oriented behavior (e.g., InstructGPT \cite{ouyang2022training}, Flan-T5 \cite{chung2024scaling}). Subsequent efforts systematized instruction datasets through large-scale task collections \cite{wang2022super}, synthetically growing the diversity and complexity of instructions \cite{xu2023wizardlm}, and small but carefully curated sets that still yield strong efficacy \cite{zhou2024lima}. For more efficient adaptation techniques, LoRA \cite{hu2021lora} and QLoRA \cite{dettmers2024qlora} enable parameter-efficient fine-tuning models through quantization and memory optimization techniques. Research continues to expand into multi-modal applications \cite{liu2024visual}, domain-specific adaptations \cite{wang2023instructuie}, and analysis of learning dynamics \cite{chung2024scaling}.

Instruction-tuning on LLMs is rapidly emerging in education, with notable applications ranging from the development of conversational AI tutors \cite{jurenka2024towards} to multimodal LLMs for art appreciation \cite{lee2024llava,lee2024llava-v2}. However, even with these growing interests, few approaches explicitly align model training with established pedagogical frameworks that structure the tutoring dialogue of the model. Therefore, this study developed a model that integrates Polya's four-step problem-solving framework as a pedagogical foundation into its behavior.

%% file: 3_method.tex
\section{Method}

\subsection{Research Procedure}
We followed an iterative cycle that linked design, data, and evaluation: literature review, Prompt-v1 design, synthetic dialogue generation from GSM8K, full fine-tuning, prompt refinement (Prompt-v2), additional data generation, re-training and evaluation. At each cycle, small batches were manually reviewed for clarity, mathematical rigor, and stage alignment. Quantitative criteria and expert-review protocols were specified before the final evaluation. 

\subsection{Problem Formulation}

Let $\mathcal{D} = {(x_i, y_i)}_{i=1}^N$ be the instruction tuning set, where $x_i$ is a math word-problem context and $y_i$ is its stage-based, step-wise solution aligned with Polya's method. We fine-tuned a pre-trained LLM $f_{\theta_0}$, initially parameterized by $\theta_0$, with a next-token objective to align model behavior to these trajectories. The instruction tuning objective can be formulated as:
$$
\mathcal{L}(\theta) = -\frac{1}{N}\sum_{i=1}^N \sum_{t=1}^{T_i} \log p_\theta(y_{i,t}|x_i, y_{i,<t})
$$
where $p_\theta(y_{i,t}|x_i, y_{i,<t})$ is the probability of generating the $t$-th token of the solution $y_i$ given the problem context $x_i$ and the previously generated tokens $y_{i,<t}$, and $T_i$ is the length of the solution $y_i$.
The fine-tuning process aims to find the optimal parameters $\theta^*$ that minimize this loss while starting from the pre-trained parameters 

$$
\theta^* = \operatorname{argmin}_\theta \mathcal{L}(\theta), \text{ initialized from } \theta_0
$$

\subsection{Data Design for Mathematical Problem-Solving}
We designed prompts with eight elements organized into three groups so that each synthetic dialogue operationalizes Polya's method and makes the assistant's turns based on phases.

To design high-quality prompts tailored for mathematical problem-solving based on Polya's method, each prompt was designed with eight elements, organized into three categories: \texttt{base elements}, \texttt{random variable elements}, and \texttt{optimized variable elements}. The \texttt{base elements} provided foundational structure, including \texttt{Situation Information} and \texttt{Utterance Guidelines}. \texttt{Situation Information} articulated the instructional situation (e.g., a teacher-student interaction during a tutoring session) and purpose in terms of Polya's method. \texttt{Utterance Guidelines} articulated communication strategies that encourage engagement, keep language accessible, and elicit clarifying or probing responses where appropriate.

To diversify contexts and discourse, the \texttt{random variable elements} varied \texttt{Student Persona} and \texttt{Math Problems}. \texttt{Student Persona} described background, strengths, and challenges in learning math, tailoring the generated data to different learner profiles. \texttt{Math Problems} were selected from GSM8K, focusing on common K-12, multi-step word problems that requires arithmetic reasoning.

The \texttt{optimized variable elements} underwent iterative refinement for a higher instruction quality and model performance. These included \texttt{Problem-Solving Stages and Flows}, \texttt{Few Shots}, \texttt{Template}, and \texttt{Instruction}. \texttt{Problem-Solving Stages and Flows} explicitly mapped to Polya’s framework with stage-specific goals, explanations, and example prompts. \texttt{Few Shots} demonstrated conversational scaffolding aligned with these stages and were formatted as JSON-like structures with \texttt{“from”: “human”} and \texttt{“from”: “gpt”} dialogue pairs. A simple \texttt{Template} defined the structure of dialogues, and the \texttt{Instruction} element guided the assistant’s behavior and scope, such as focusing only on “Understanding the Problem.”

\subsection{Data Generation}
Using the prompt schema above, GPT-4o was prompted with the math problems sampled from GSM8K dataset to produce Polya-aligned tutoring dialogues \cite{cobbe2021gsm8k}. Researchers reviewed samples for clarity, correctness, and instructional appropriateness of the dialogues. This process was iterative and adaptive, enabling refinement of both prompts and dialogue structures based on the quality and patterns of the model’s responses.

\subsection{Instruction Tuning}

The instruction tuning strategies focused on instruction tuning using a ChatML template format \cite{huggingface_chat_templating}. Let $\mathcal{D}_{\text{inst}} = {(c_i, r_i)}_{i=1}^M$ be our instruction tuning dataset, where $c_i$ represents the context (including the mathematical problem and any prior conversation) and $r_i$ is the corresponding assistant's response.
The instruction tuning objective can be formulated as:
$$
\mathcal{L}{\text{inst}}(\theta) = -\frac{1}{M}\sum{i=1}^M \sum_{t=1}^{T_i} \log p_\theta(r_{i,t}|c_i, r_{i,<t})
$$
where $p_\theta(r_{i,t}|c_i, r_{i,<t})$ is the probability of generating the $t$-th token of the response $r_i$ given the context $c_i$ and the previously generated tokens $r_{i,<t}$, and $T_i$ is the length of the response $r_i$.

The ChatML template was used to structure the input, with each example formatted using \texttt{<|im\_start|>}  and \texttt{<|im\_end|>}  tags to delineate human and assistant parts. The model was trained exclusively on the assistant's part, excluding the system and user parts.
Full fine-tuning was performed without using Parameter-Efficient Fine-Tuning (PEFT) techniques. This approach allowed for comprehensive updates to all model parameters, potentially leading to better performance at the cost of higher computational requirements. The training was conducted for a single epoch over the entire dataset $\mathcal{D}_{\text{inst}}$, balancing between model improvement and computational efficiency.

%% file: 4_result.tex
\section{Result}

\subsection{Experiment Settings}

\subsubsection{Model}

In our experiments, we employed several variations of the LLaMA 3.1-8B model \cite{meta_llama3_1} to evaluate the effectiveness of our approach in different mathematical domains. All models share the same architecture but differ in training objectives and fine-tuning datasets.

Specifically, we included the following variants: (1) the \texttt{base} model, which is pretrained but not instruction-tuned; (2) the \texttt{instruct} model, which is tuned for general-purpose instruction following, the ability of an LLM to follow user instructions; (3) \texttt{polya-v2}, a fine-tuned model on a curated math corpus focused on diverse problem types; (4) \texttt{metamath}, trained on a formal mathematical reasoning dataset; and (5) \texttt{metamath + polya-v2}, which combines the strengths of both by sequential fine-tuning. The whole list of models is summarized in Table~\ref{tab:model_variants}.

\begin{table}[h]
\centering
\caption{LLaMA 3.1-8B model variants used}
\label{tab:model_variants}
\resizebox{\textwidth}{!}{%
\begin{tabular}{ll}
\toprule
\textbf{Model} & \textbf{Description} \\
\midrule
\texttt{base} & Pretrained only, no instruction tuning \\
\texttt{instruct} & Instruction-tuned variant from Meta \\
\texttt{polya-v2} & Finetuned on the Polya v2 math corpus \\
\texttt{metamath} & Finetuned on the Metamath dataset \\
\texttt{metamath + polya-v2} & Sequentially tuned on Metamath then Polya v2 \\
\bottomrule
\end{tabular}
}
\end{table}

We evaluated each model in tasks from three domains: \textit{Arithmetic}, \textit{Measurement}, and \textit{Geometry}.

\subsubsection{Dataset}

The dataset consisted of approximately 32,000 samples of dialogue data. These samples were generated following the structured prompt design and data generation procedures described in the \textit{Data Design for Mathematics Problem-Solving} and \textit{Data Generation} sections.

\begin{figure*}[t]
    \centering
    \includegraphics[width=\textwidth]{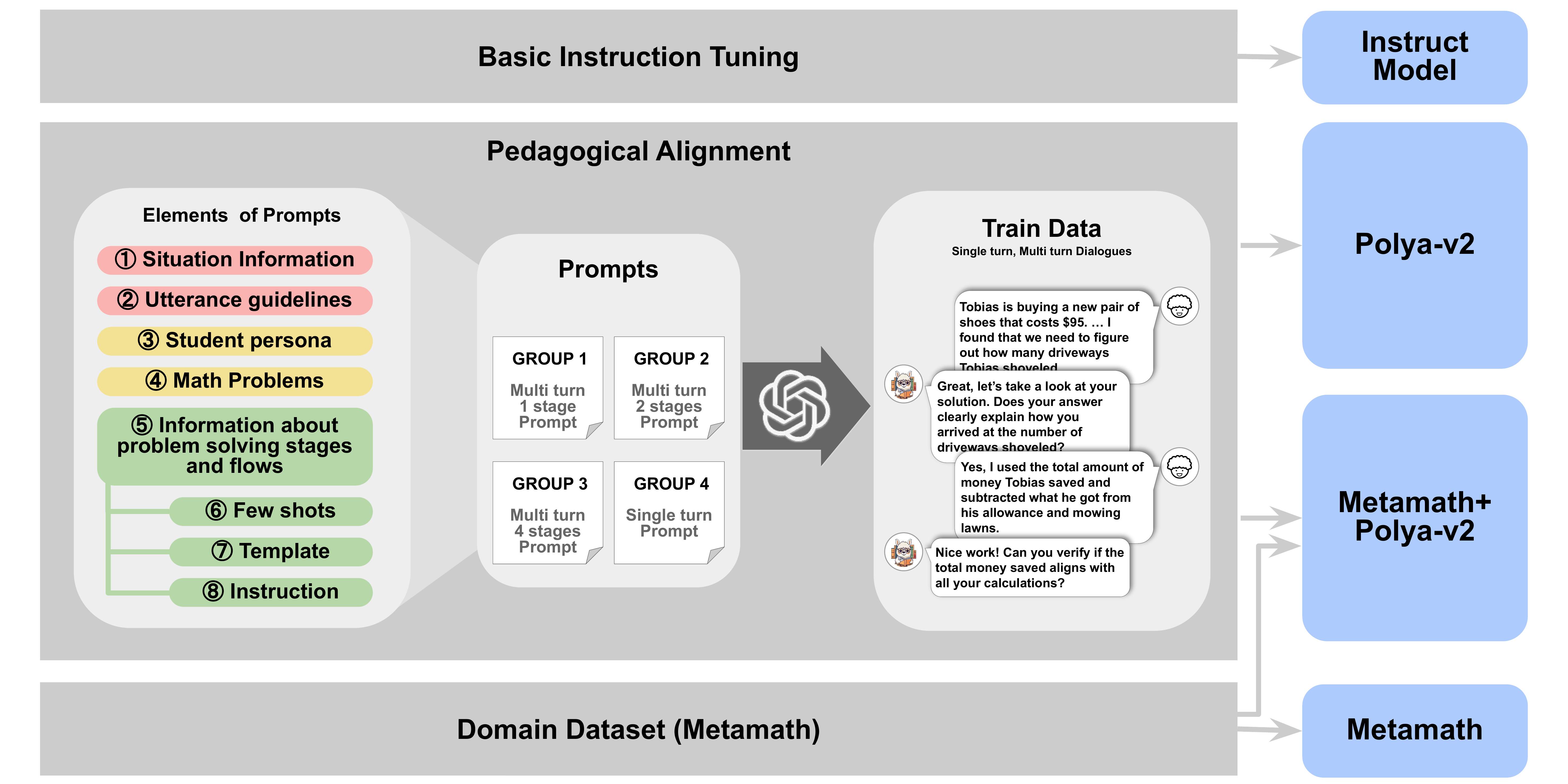}
    \caption{Process of train dataset generation. Each stage links to Polya's critical stages.}
    \label{fig:data}
\end{figure*}

\subsubsection{Training Configuration}

The Axolotl \cite{axolotl2025} was utilized for training. The training was conducted with a batch size of 1, using eight A100 GPUs. The learning rate was set at 0.0002, with 100 warm-up steps and a weight decay of 0.1. Additionally, DeepSpeed's ZeRO-2 \cite{deepspeed2025} optimization was employed to efficiently manage the computational load. This configuration ensured that the training process was both effective and resource-efficient, leveraging state-of-the-art hardware and optimization techniques.

\subsubsection{Model Evaluation}

The model evaluation was conducted in two stages: Polya stage annotation by researchers and expert validation. The researchers engaged in 10–20 turn dialogues with each model using problems from three categories; Arithmetic, Measurement, and Geometry, and collected conversational data for analysis.

Each dialogue was manually annotated according to Polya's four problem-solving stages: \textit{Understanding the problem}, \textit{Devising a plan}, \textit{Carrying out the plan}, and \textit{Looking back}. Based on these annotations, the researchers conducted a structured quantitative assessment of the reasoning behavior of each model.

To validate these findings, domain experts reviewed selected samples. They completed a survey consisting of both Likert-scale questions (quantitative data) and open-ended responses (qualitative data). The ratings were statistically analyzed, and the written feedback was coded to identify recurring themes regarding the strengths and weaknesses of each model.

\subsection{Evaluation result}

\subsubsection{Polya stage annotation results}

The annotation results reveal clear distinctions in model behaviors when analyzed through the lens of Polya's problem-solving framework. Notably, the two models trained on math-specific data, \texttt{polya-v2} and \texttt{metamath}, exhibited the most balanced stage distributions across three math domains (Arithmetic, Measurement, and Geometry). These models consistently guided learners from the initial stage of understanding the problem to more advanced stages involving higher-order reasoning. Furthermore, they showed the lowest error rates among all models (\texttt{polya-v2}: 0.0\%, \texttt{metamath}: 0.4\%).

In contrast, models developed for general-purpose use, \texttt{base} and \texttt{instruct}, exhibited higher error rates and lower performance levels. A common pattern of errors observed in these models was to bypass Polya's structured problem-solving stages. Rather than offering stage-appropriate scaffolding in response to learner queries, they sometimes provided direct answers prematurely or completed solution procedures in a single conversational turn. This approach curtailed opportunities for learners to engage in self-directed exploration, suggesting that generic instruction tuning is not sufficient to provide pedagogical tutoring in LLMs.

Interestingly, the combined model \texttt{Metamath + polya-v2}, which sequentially fine-tuned with math-specific and pedagogical datasets, also exhibited high error rates. This result may suggest potential conflicts or interference between domain-specific knowledge and pedagogically oriented reasoning. Figure 2 and Table II present details of the normalized stage distributions for each model. 

\begin{figure}[t]
    \centering
    \includegraphics[width=0.5\textwidth]{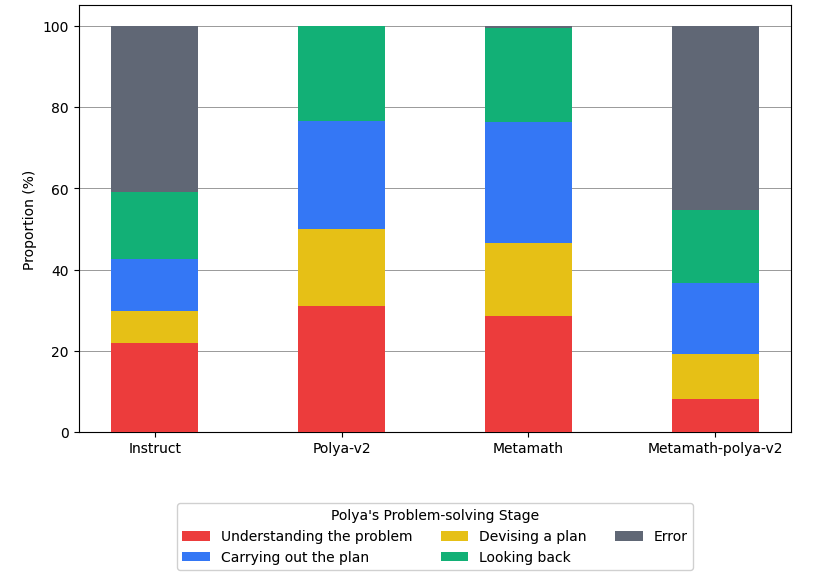}
    \caption{Normalized Polya-stage distributions across models}
    \label{fig:polya_stage}
\end{figure}

\begin{table}[htbp]
\caption{Summary of Model Evaluation by Polya Stage}
\centering
\resizebox{\columnwidth}{!}{%
\begin{tabular}{llrlllll}
\toprule
Model & Math Domain & \shortstack{Average \\ Conv. Length} & Stage 1 & Stage 2 & Stage 3 & Stage 4 & Error rate \\
\midrule
Base & Arithmetic & 3.7 & 0.0\% & 0.0\% & 0.0\% & 0.0\% & 100.0\% \\
 & Measurement & 2.7 & 0.0\% & 0.0\% & 0.0\% & 0.0\% & 100.0\% \\
 & Geometry & 2.5 & 0.0\% & 0.0\% & 0.0\% & 0.0\% & 100.0\% \\
 & Average & 3.0 & 0.0\% & 0.0\% & 0.0\% & 0.0\% & 100.0\% \\
Instruct & Arithmetic & 9.7 & 23.3\% & 0.0\% & 14.8\% & 19.0\% & 42.9\% \\
 & Measurement & 9.1 & 31.4\% & 14.0\% & 14.6\% & 13.7\% & 26.3\% \\
 & Geometry & 9.4 & 10.9\% & 9.3\% & 9.3\% & 16.9\% & 53.7\% \\
 & Average & 9.4 & 21.9\% & 7.8\% & 12.9\% & 16.5\% & 41.0\% \\
polya-v2 & Arithmetic & 19.4 & 32.4\% & 18.1\% & 31.8\% & 17.7\% & 0.0\% \\
 & Measurement & 16.9 & 28.7\% & 16.7\% & 25.9\% & 28.7\% & 0.0\% \\
 & Geometry & 17.7 & 31.8\% & 22.0\% & 22.4\% & 23.8\% & 0.0\% \\
 & Average & 18.0 & 31.0\% & 18.9\% & 26.7\% & 23.4\% & 0.0\% \\
metamath & Arithmetic & 19.7 & 25.0\% & 19.8\% & 38.8\% & 15.1\% & 1.3\% \\
 & Measurement & 18.0 & 35.0\% & 15.5\% & 17.9\% & 31.5\% & 0.0\% \\
 & Geometry & 19.7 & 25.6\% & 18.6\% & 32.7\% & 23.1\% & 0.0\% \\
 & Average & 19.1 & 28.5\% & 17.9\% & 29.8\% & 23.2\% & 0.4\% \\
\bottomrule
\end{tabular}%
}
\label{tab:model-polya}
\end{table}

Notable differences in model behavior according to the mathematical domains were also observed. In arithmetic tasks, which primarily require the application of given numerical values, model responses tended to concentrate on the execution (Stage 3). However, in measurement tasks, model responses contained a relatively higher proportion of Stage 1 and Stage 4. This may reflect the cognitive demands of the tasks, such as interpreting contextual information and verifying reasoning in unit conversions. Models exhibited the most evenly distributed stage engagement in geometry tasks, which require a comprehensive understanding and reasoning due to tasks' inferential nature. These variations indicate that certain mathematical domains may better prompt specific reasoning behaviors depending on task structure and complexity.

\subsubsection{Expert Validation Results}

Expert evaluation was conducted using a mixed-methods approach, combining Likert-scale ratings and open-ended feedback from six mathematics education specialists. Quantitatively, experts generally rated the chatbot positively in terms of structure, tone, and alignment with educational intentions. The highest-scoring items were those related to self-level feedback, with an average score of 4.22 on a 5-point scale. The chatbot was also well-received for its respectful and encouraging tone toward students. However, the lowest ratings were given to task-level feedback, such as explaining why the answer is correct or not, typically around 2.8 to 3.2. 

Narrative feedback clarified why these patterns emerged. Reviewers noted that the chatbot’s frequent meta-cognitive prompts helped students monitor and review their own problem-solving, thereby sustaining engagement and self-regulation. Also, its rich, encouraging language also generated a welcome sense of recognition that many believed could foster more positive attitudes toward mathematics, while the conversational format let students pose questions and structure solutions at their own pace.

Yet the same reviewers stressed several shortcomings. First, they argued that deeper reflective planning is required: prompts should guide students to articulate what they do not yet understand and to map out next steps, rather than simply flagging errors. Second, they observed that feedback often felt generic; indiscriminate praise such as “Great job!” did little to acknowledge the specific thinking processes of learners with diverse backgrounds or skill levels. Third, mathematical rigor occasionally lagged, with misconceptions left uncorrected or addressed only superficially. Finally, although the chatbot sometimes invoked Polya’s “Looking Back” stage, it rarely redirected students to earlier reasoning when conceptual gaps persisted, limiting the depth of reflection that took place.

Across the board, reviewers emphasized that skilled human teachers remain essential to interpret contextual subtleties, including non-verbal cues, classroom dynamics, and moment-to-moment affect, and to adapt feedback in real time. They recommended that large-language-model systems move toward this flexibility by leveraging extensive paired question-and-answer datasets and by integrating real-time signals about student affect and performance.

In sum, the panel concluded that the chatbot shows considerable promise as a structured, motivational, and fatigue-free support tool, but that substantial gains could still be made by strengthening its mathematical rigor, personalizing its feedback, and embedding more robust scaffolds that help students plan reflectively and revisit earlier reasoning stages when necessary.

%% file: 5_discussion.tex
\section{Discussion}

This study examined how instruction tuning grounded in Polya’s problem-solving framework can reshape the role of conversational LLMs in mathematics education. Rather than serving merely as answer generators, such models can act as pedagogical partners that scaffold students’ reasoning and foster metacognitive engagement. The below synthesizes the main findings, highlights limitations, and outlines broader implications for research and practice.

First, pedagogical alignment guided students through the stages of understanding, planning, execution, and review, thereby clarifying cognitive structures and encouraging reflective responses. Notably, in the geometry domain, learners produced reflective utterances regardless of correctness, suggesting an enhancement of metacognitive regulation. This aligns with prior research linking metacognitive strategies to mathematical achievement \cite{garofalo1985metacognition,schoenfeld2014mathematical}. Recent empirical studies also support the claim that AI tutoring fosters deeper cognitive engagement and, in some cases, surpasses traditional active-learning formats \cite{kestin2025ai,wang2024tutor}.

Second, the model exhibited limitations in personalization and mathematical rigor. It failed to sufficiently account for differences in learners’ prior knowledge or problem-solving strategies, and in some cases reinforced misconceptions. For example, in algebra tasks the model occasionally strengthened erroneous procedures, thereby highlighting risks of automation bias \cite{maclennan1996survey}. Moreover, when sequentially fine-tuned on Metamath and subsequently on Polya-v2, the model underperformed compared with single-dataset tuning, reflecting catastrophic forgetting \cite{french1999catastrophic,goodfellow2014generative,luo2025empirical}. Meta-analyses further confirm that the effectiveness of AI in education depends critically on adaptive feedback and personalization \cite{dong2025examining,letourneau2025systematic}.

Third, compared with rule-based tutoring systems or general instruction-tuned models, our approach preserved exploratory thinking and learner autonomy. While rule-based systems emphasize correctness, Llama-Polya scaffolded stepwise reasoning and elicited inquiry-oriented dialogue. This contrasts with earlier findings that excessive feedback may reduce learner independence \cite{hattie2007power,kalyuga2007expertise}. Nevertheless, excessive scaffolding risks inducing cognitive overload, which must be carefully managed \cite{sweller1988load}.

Fourth, Polya’s framework can be situated within contemporary learning theories. For instance, self-regulated learning emphasizes iterative goal setting, strategy use, and self-evaluation \cite{zimmerman2002becoming}, and the model’s structured prompts may support these processes. From the perspective of distributed cognition \cite{hutchins1995cognition}, teachers and AI systems can co-construct learning communities by complementing one another’s strengths. Similarly, knowledge-building theory highlights collective meaning-making \cite{scardamalia2006knowledge}, which AI can support through iterative scaffolding. At the same time, responsible integration requires robust evaluation frameworks \cite{jurenka2024towards} and sensitivity to classroom realities such as teacher workload, learner trust, and accessibility \cite{luckin2016intelligence}. Recent evidence also warns that unregulated use of generative AI may undermine learning outcomes \cite{bastani2025generative}.

Finally, future research should adopt more empirical and practice-oriented directions. Adaptive prompting strategies \cite{cai2024power} are needed to tailor interactions to individual learner profiles. Mixed-method classroom studies---including observational, analytic, and interview-based approaches---should be employed to evaluate AI integration in authentic contexts. Teacher training programs could also incorporate AI-based simulations, allowing teachers to practice instructional strategies across diverse learner scenarios. Such efforts align with growing evidence that AI tutoring can outperform traditional instruction \cite{kestin2025ai} while mitigating risks associated with unguarded deployment \cite{bastani2025generative}.

In summary, Polya-based instruction tuning demonstrates strong potential to transform LLMs into pedagogical partners that promote reflective reasoning and learner autonomy. Nonetheless, challenges remain in personalization, mathematical rigor, and dataset interference. Addressing these issues requires methodological advances in adaptive modeling and rigorous validation in classroom contexts.

%% file: 6_conclusion.tex
\section{Conclusion}

This study introduced LLaMA-Polya, an instruction-tuned language model aligned with Polya’s four-step method to scaffold mathematical problem-solving in dialogue. Through both researcher annotation and expert validation, we found that domain-specific fine-tuning significantly improved the model’s ability to provide stage-aware scaffolding during tutoring exchanges. However, challenges remain in terms of personalization, robust error correction, and dynamic instructional responsiveness.

Our findings underscore the importance of pedagogical grounding in LLM development for educational purpose. Instruction tuning anchored in learning/instructional theories and principles theory---rather than task format alone---can enable models to better simulate expert-like teaching behavior. Future work should explore multi-turn adaptive prompting, curriculum-aware fine-tuning, and hybrid human-AI collaboration strategies to further enhance the effectiveness of AI in educational settings.

\section{Generative AI Usage Disclosure}

In accordance with IEEE Guidelines for Generative AI Usage (IEEE Robotics and Automation Society), the authors disclose the use of generative artificial intelligence (AI) tools in this manuscript. We employed an AI tool for language polishing, grammar correction, and stylistic improvement throughout the writing process. No AI tool was used for formulating the logic, generating results, deriving interpretations, or making theoretical claims. All analytical reasoning, experiment design, data interpretation, and conclusions are solely the work of the authors.

%% file: 7_appendix_survey.tex
\clearpage
\onecolumn
\appendix

\begin{center}
{\large \textbf{SURVEY QUESTIONNAIRE}}\\[0.5em]
\end{center}

\vspace{0.5cm}

\begin{center}
\begin{longtable}{|p{0.08\textwidth}|p{0.78\textwidth}|p{0.08\textwidth}|}
\caption{Survey Questionnaire Items} \\
\hline
\textbf{Num} & \textbf{Item} & \textbf{Scale} \\
\hline
\endfirsthead

\multicolumn{3}{c}%
{{\bfseries Table \thetable{} -- continued from previous page}} \\
\hline
\textbf{Num} & \textbf{Item} & \textbf{Scale} \\
\hline
\endhead

\hline \multicolumn{3}{|r|}{{Continued on next page}} \\ \hline
\endfoot

\hline
\endlastfoot

\multicolumn{3}{|c|}{\textbf{[Design Principles]}} \\
\hline

\multicolumn{3}{|l|}{\textbf{Section 1: Contextual Alignment}} \\
\multicolumn{3}{|p{0.94\textwidth}|}{\footnotesize In this section, we evaluate whether LLaMA-Polya (an LLM-based chatbot) can be used in the context of the tool's intended use as designed.} \\
\hline

1 & Situation: The conversation must simulate a realistic classroom or tutoring environment where a teacher and student engage to solve a math problem. & 1-5 \\
\hline

2 & Purpose: The goal of the tool is to help the student work through each stage of the problem-solving process, promoting understanding and active learning. & 1-5 \\
\hline

3 & Role: The tool acts as a guide and mentor, while the student adopts the role of a curious learner. & 1-5 \\
\hline

4 & Constraints: Ensure that the flow aligns with Polya's 4-step problem-solving stages and follows the specific case instructions (\textit{e.g. difficulties in specific steps}). & 1-5 \\
\hline

\multicolumn{3}{|l|}{\textbf{Section 2: LLM-based Chatbot Utterances}} \\
\multicolumn{3}{|p{0.94\textwidth}|}{\footnotesize In this section, we evaluate whether LLaMA-Polya (an LLM-based chatbot) conforms to the style, attitude, and content of the utterances intended by the design.} \\
\hline

1 & Use simplified language \& Keep respect and patience

\textit{e.g.}

\textit{LLM-based Chatbot: Great job! Can you simplify and double-check?
(Student: 8+6 = 14.)}

\textit{LLM-based Chatbot: Excellent work! Now, let's look back. Does this perimeter value that we found out make sense based on the dimensions given in the problem?} & 1-5 \\
\hline

2 & Support students to maintain focus, Encourage engagement

\textit{e.g.}

\textit{LLM-based Chatbot: Hold on, we are almost there. Think of details like the cost of the adult tickets, child tickets, and how many of each. What would the next step be?}

\textit{(Student: Two adult tickets are 51*2 = \$102. Two child tickets are 25.5*2=\$51. Then the final price is \$153!)}

\textit{LLM-based Chatbot: Excellent! Now let's reflect on our solution. Does this total make sense given the ticket prices and quantities?} & 1-5 \\
\hline

3 & LLM-based chatbots use feedback strategies that ask for re-explanation or clarification to help students actively construct knowledge.

\textit{e.g.}
\begin{itemize}
\item Encourage exploration
\item Use Clarifying Questions: Rephrase, Clarify, Elaborate
\item Avoid accepting incorrect answers \& Handle "I Don't Know" Responses
\end{itemize} & 1-5 \\
\hline

\multicolumn{3}{|l|}{\textbf{Section 3: Student personas and dialog adapted accordingly}} \\
\multicolumn{3}{|p{0.94\textwidth}|}{\footnotesize In this section, we evaluate whether LLaMA-Polya (an LLM-based chatbot) adjusts the utterances according to the characteristics of learners as intended by the design.} \\
\hline

1 & Adapt responses considering different student characteristics and cultural backgrounds (\textit{e.g., age, nationality, mathematical ability}) & 1-5 \\
\hline

2 & Respond properly considering the situation when the student exhibits curiosity and occasionally struggles with one or more steps in the problem-solving process (based on the case scenarios).

\textbullet\ Explanation of difficult notions, Clarification of questions, e.t.c.

\textit{e.g.}

\textit{(Student: I don't know. Please explain easily.) }

\textit{LLM-based Chatbot: Sure, let's simplify. We need to find the perimeter of Maria's garden. What are the key pieces of information given in the problem?} & 1-5 \\
\hline

3 & Respond properly to the students' requests for clarification, or logical reasoning demonstrations. & 1-5 \\
\hline

\multicolumn{3}{|l|}{\textbf{Section 4: Polya's Problem-Solving Stages}} \\
\multicolumn{3}{|p{0.94\textwidth}|}{\footnotesize In this section, we evaluate whether LLaMA-Polya (an LLM-based chatbot) provides utterances to learners according to Polya's Problem-Solving Stages} \\
\hline

1 & Follow Polya's problem-solving stages in the right order. & 1-5 \\
\hline

2 & Provide detailed explanation in a specific stage when students request. & 1-5 \\
\hline

3 & If a student struggles with one or more steps in the problem-solving process, the tool goes back to the former step in order to learn more. & 1-5 \\
\hline

\multicolumn{3}{|c|}{\textbf{[Pedagogical Validity]}} \\
\hline

\multicolumn{3}{|l|}{\textbf{Section 1: Task-level Feedback}} \\
\multicolumn{3}{|p{0.94\textwidth}|}{\footnotesize Task-level feedback concerns whether the chatbot feedback effectively informs students about the quality or correctness of their answers, including both final and intermediate responses during problem-solving.} \\
\hline

1 & The LLM feedback effectively evaluates the correctness or quality of a student's answer. & 1-5 \\
\hline

2 & The LLM feedback provides clear explanations for why a student's answer is correct or incorrect. & 1-5 \\
\hline

\multicolumn{3}{|l|}{\textbf{Section 2: Process-level Feedback}} \\
\multicolumn{3}{|p{0.94\textwidth}|}{\footnotesize This section evaluates whether the LLM feedback effectively provides information on the task-solving process used to attain a result or answer, including identifying errors and suggesting ways to correct them.} \\
\hline

1 & The LLM feedback accurately identifies and indicates location of specific errors or (mis) understandings in the student's problem-solving process. & 1-5 \\
\hline

2 & The LLM feedback provides descriptions or explanations of errors that students make. & 1-5 \\
\hline

3 & The LLM feedback offers helpful (actionable?) hints or guidance to students on how to proceed in task solving. & 1-5 \\
\hline

\multicolumn{3}{|l|}{\textbf{Section 3: Self-regulation level Feedback}} \\
\multicolumn{3}{|p{0.94\textwidth}|}{\footnotesize This section evaluates whether the LLM feedback helps students monitor, adjust, and regulate their use of task-solving methods or learning strategies in their learning and task-solving processes.} \\
\hline

1 & The LLM feedback includes prompts or metacognitive questions encouraging students to self-evaluate or explain their answers. & 1-5 \\
\hline

2 & The LLM feedback includes prompts or metacognitive questions on parts of students' problem-solving processes, or solution strategy (\textit{e.g., asking students to explain their steps or approaches}). & 1-5 \\
\hline

\multicolumn{3}{|l|}{\textbf{Section 4: Self-level Feedback}} \\
\multicolumn{3}{|p{0.94\textwidth}|}{\footnotesize This section evaluates whether the LLM feedback provides general positive or negative evaluations of students' work, effort, or personal characteristics.} \\
\hline

1 & The LLM feedback effectively acknowledges and provides general evaluations of students' work, whether positive or negative (\textit{e.g., "Great work!"}) & 1-5 \\
\hline

2 & The LLM feedback appropriately recognizes and encourages students' efforts during the task (\textit{e.g., "You made a good effort during counting!"}) & 1-5 \\
\hline

3 & The LLM feedback recognizes and provides positive or negative evaluations of students' personal characteristics (\textit{e.g., "You are good at counting numbers!"}) & 1-5 \\
\hline

\multicolumn{3}{|l|}{\textbf{Section 5: Open-ended Questions}} \\
\hline

1 & In your opinion, what are the key strengths of the LLM-based agent's feedback at each level (task-level, process-level, self-regulation level, or self-level)? & \\
\hline

2 & Are there any aspects where the LLM-based agent's feedback could be improved to better align with pedagogical best practices? & \\
\hline

3 & Based on your experience, how would you compare the feedback provided by the LLM model with that of a skilled human educator? & \\
\hline

\end{longtable}
\end{center}

\addcontentsline{toc}{section}{Appendix: Survey Questionnaire}

%% file: main.bbl
\begin{thebibliography}{69}

\bibitem{axolotl2025}
Axolotl AI Cloud: Axolotl: A Tool for Streamlined Post-Training of AI Models. GitHub repository (2025), \url{https://github.com/axolotl-ai-cloud/axolotl}

\bibitem{baidoo2023education}
Baidoo-Anu, D., Ansah, L.O.: Education in the era of generative artificial intelligence (AI): Understanding the potential benefits of ChatGPT in promoting teaching and learning. Journal of AI \textbf{7}(1), 52--62 (2023)

\bibitem{bastani2025generative}
Bastani, H., Bastani, O., Sungu, A., Ge, H., Kabakc{\i}, {\"O}., Mariman, R.: Generative AI without guardrails can harm learning: Evidence from high school mathematics. Proceedings of the National Academy of Sciences \textbf{122}(26), e2422633122 (2025)

\bibitem{bray2017technology}
Bray, A., Tangney, B.: Technology usage in mathematics education research--A systematic review of recent trends. Computers \& Education \textbf{114}, 255--273 (2017)

\bibitem{cai2024power}
Cai, S., Mensah-Boateng, T., Kuksov, X., Yuan, J., Tang, S.: The Power of Adaptation: Boosting In-Context Learning through Adaptive Prompting. arXiv preprint arXiv:2412.17891 (2024)

\bibitem{casalino2021deep}
Casalino, G., Grilli, L., Limone, P., Santoro, D., Schicchi, D.: Deep learning for knowledge tracing in learning analytics: an overview. TeleXbe (2021)

\bibitem{chang2006computer}
Chang, K.-E., Sung, Y.-T., Lin, S.-F.: Computer-assisted learning for mathematical problem solving. Computers \& Education \textbf{46}(2), 140--151 (2006)

\bibitem{christ2024mathwell}
Christ, B.R., Kropko, J., Hartvigsen, T.: MATHWELL: Generating Educational Math Word Problems at Scale. arXiv preprint arXiv:2402.15861 (2024)

\bibitem{chung2024scaling}
Chung, H.W., Hou, L., Longpre, S., Zoph, B., Tay, Y., Fedus, W., Li, Y., Wang, X., Dehghani, M., Brahma, S., et al.: Scaling instruction-finetuned language models. Journal of Machine Learning Research \textbf{25}(70), 1--53 (2024)

\bibitem{cobbe2021gsm8k}
Cobbe, K., Kosaraju, V., Bavarian, M., Chen, M., Jun, H., Kaiser, L., Plappert, M., Tworek, J., Hilton, J., Nakano, R., Hesse, C., Schulman, J.: Training Verifiers to Solve Math Word Problems. arXiv preprint arXiv:2110.14168 (2021)

\bibitem{cunska2020effective}
Cunska, A., et al.: Effective learning strategies and Artificial Intelligence (AI) support for accelerated math acquisition. In: European Proceedings of International Conference on Education and Educational Psychology. European Publisher (2020)

\bibitem{dawson2023technology}
Dawson, P., Henderson, M., Ryan, T., Mahoney, P., Boud, D., Phillips, M., Molloy, E.: Technology and feedback design. In: Learning, Design, and Technology: An International Compendium of Theory, Research, Practice, and Policy, pp.~695--739. Springer (2023)

\bibitem{deepspeed2025}
Microsoft: DeepSpeed: A Deep Learning Optimization Library. GitHub repository (2025), \url{https://github.com/deepspeedai/DeepSpeed}

\bibitem{dettmers2024qlora}
Dettmers, T., Pagnoni, A., Holtzman, A., Zettlemoyer, L.: QLoRA: Efficient finetuning of quantized LLMs. Advances in Neural Information Processing Systems \textbf{36} (2024)

\bibitem{dong2025examining}
Dong, L., Tang, X., Wang, X.: Examining the effect of artificial intelligence in relation to students' academic achievement in classroom: A meta-analysis. Computers and Education: Artificial Intelligence, 100400 (2025)

\bibitem{drijvers2010introduction}
Drijvers, P., Mariotti, M.-A., Olive, J., Sacrist{\'a}n, A.I.: Introduction to section 2. Mathematics Education and Technology-Rethinking the Terrain: The 17th ICMI Study, pp.~81--87. Springer (2010)

\bibitem{duval2006cognitive}
Duval, R.: A cognitive analysis of problems of comprehension in a learning of mathematics. Educational Studies in Mathematics \textbf{61}, 103--131 (2006)

\bibitem{feng2024exploring}
Feng, W., Lee, J., McNichols, H., Scarlatos, A., Smith, D., Woodhead, S., Ornelas, N.O., Lan, A.: Exploring automated distractor generation for math multiple-choice questions via large language models. arXiv preprint arXiv:2404.02124 (2024)

\bibitem{french1999catastrophic}
French, R.M.: Catastrophic forgetting in connectionist networks. Trends in Cognitive Sciences \textbf{3}(4), 128--135 (1999)

\bibitem{garofalo1985metacognition}
Garofalo, J., Lester, F.K.: Metacognition, cognitive monitoring, and mathematical performance. Journal for Research in Mathematics Education \textbf{16}(3), 163--176 (1985)

\bibitem{geiger2012technology}
Geiger, V., Forgasz, H., Tan, H., Calder, N., Hill, J.: Technology in mathematics education. In: Research in Mathematics Education in Australasia 2008-2011, pp.~111--141. Brill (2012)

\bibitem{goldin1997chapter}
Goldin, G.A.: Chapter 4: Observing mathematical problem solving through task-based interviews. Journal for Research in Mathematics Education. Monograph, 40--177 (1997)

\bibitem{goodfellow2014generative}
Goodfellow, I.J., Pouget-Abadie, J., Mirza, M., Xu, B., Warde-Farley, D., Ozair, S., Courville, A., Bengio, Y.: Generative adversarial nets. Advances in Neural Information Processing Systems \textbf{27} (2014)

\bibitem{haddad2002technologies}
Haddad, W.D., Draxler, A.: Technologies for education: Potential, parameters, and prospects. Washington, DC: AED \textbf{20}, 2006 (2002)

\bibitem{hattie2007power}
Hattie, J., Timperley, H.: The power of feedback. Review of Educational Research \textbf{77}(1), 81--112 (2007)

\bibitem{henderson1953problem}
Henderson, K.B., Pingry, R.E.: Problem solving in mathematics. The Learning of Mathematics: Its Theory and Practice, 228--270 (1953)

\bibitem{holmes2023artificial}
Holmes, W., Bialik, M., Fadel, C.: Artificial Intelligence in Education. Globethics Publications (2023)

\bibitem{hu2021lora}
Hu, E.J., Shen, Y., Wallis, P., Allen-Zhu, Z., Li, Y., Wang, S., Wang, L., Chen, W.: LoRA: Low-rank adaptation of large language models. arXiv preprint arXiv:2106.09685 (2021)

\bibitem{hu2024teaching}
Hu, B., Zheng, L., Zhu, J., Ding, L., Wang, Y., Gu, X.: Teaching Plan Generation and Evaluation With GPT-4: Unleashing the Potential of LLM in Instructional Design. IEEE Transactions on Learning Technologies (2024)

\bibitem{huggingface_chat_templating}
Hugging Face: Chat Templates (2025), \url{https://huggingface.co/docs/transformers/en/chat_templating}

\bibitem{hutchins1995cognition}
Hutchins, E.: Cognition in the Wild. MIT Press (1995)

\bibitem{hwang2021roles}
Hwang, G.-J., Tu, Y.-F.: Roles and research trends of artificial intelligence in mathematics education: A bibliometric mapping analysis and systematic review. Mathematics \textbf{9}(6), 584 (2021)

\bibitem{jurenka2024towards}
Jurenka, I., Kunesch, M., McKee, K., et al.: Towards Responsible Development of Generative AI for Education: An Evaluation-Driven Approach. Preprint (2024), \url{https://goo.gle/LearnLM}

\bibitem{kalyuga2007expertise}
Kalyuga, S.: Expertise reversal effect and its implications for learner-tailored instruction. Educational Psychology Review \textbf{19}(4), 509--539 (2007)

\bibitem{kaput2020technology}
Kaput, J., Hegedus, S., Lesh, R.: Technology becoming infrastructural in mathematics education. In: Foundations for the Future in Mathematics Education, pp.~173--191. Routledge (2020)

\bibitem{kasneci2023chatgpt}
Kasneci, E., Se{\ss}ler, K., K{\"u}chemann, S., Bannert, M., Dementieva, D., Fischer, F., Gasser, U., Groh, G., G{\"u}nnemann, S., H{\"u}llermeier, E., et al.: ChatGPT for good? On opportunities and challenges of large language models for education. Learning and Individual Differences \textbf{103}, 102274 (2023)

\bibitem{kestin2025ai}
Kestin, G., Miller, K., Klales, A., Milbourne, T., Ponti, G.: AI tutoring outperforms in-class active learning: an RCT introducing a novel research-based design in an authentic educational setting. Scientific Reports \textbf{15}(1), 17458 (2025)

\bibitem{kim2022learning}
Kim, J., Lee, H., Cho, Y.H.: Learning design to support student-AI collaboration: Perspectives of leading teachers for AI in education. Education and Information Technologies \textbf{27}(5), 6069--6104 (2022)

\bibitem{kumar2023math}
Kumar, H., Rothschild, D.M., Goldstein, D.G., Hofman, J.M.: Math Education with Large Language Models: Peril or Promise? Available at SSRN 4641653 (2023)

\bibitem{lee2016appropriate}
Lee, C.I.: An appropriate prompts system based on the Polya method for mathematical problem-solving. Eurasia Journal of Mathematics, Science and Technology Education \textbf{13}(3), 893--910 (2016)

\bibitem{lee2024llava}
Lee, U., Jeon, M., Lee, Y., Byun, G., Son, Y., Shin, J., Ko, H., Kim, H.: LLaVA-Docent: Instruction Tuning with Multimodal Large Language Model to Support Art Appreciation Education. arXiv preprint arXiv:2402.06264 (2024)

\bibitem{lee2024llava-v2}
Lee, U., Son, Y., Shin, J., Byun, G., Lee, Y., Koh, J., Jeon, M., Kim, H.: LLaVA-Docent-V2: Improving Data Quality to Train Large Multimodal Models for Art Appreciation Education. Preprint on ResearchGate (2024)

\bibitem{letourneau2025systematic}
L{\'e}tourneau, A., Deslandes Martineau, M., Charland, P., Karran, J.A., Boasen, J., L{\'e}ger, P.M.: A systematic review of AI-driven intelligent tutoring systems (ITS) in K-12 education. npj Science of Learning \textbf{10}(1), 29 (2025)

\bibitem{liu2024visual}
Liu, H., Li, C., Wu, Q., Lee, Y.J.: Visual instruction tuning. Advances in Neural Information Processing Systems \textbf{36} (2024)

\bibitem{luckin2016intelligence}
Luckin, R., Holmes, W.: Intelligence unleashed: An argument for AI in education. Tech. rep., UCL Knowledge Lab (2016)

\bibitem{luckin2017towards}
Luckin, R.: Towards artificial intelligence-based assessment systems. Nature Human Behaviour \textbf{1}(3), 0028 (2017)

\bibitem{luo2025empirical}
Luo, Y., Yang, Z., Meng, F., Li, Y., Zhou, J., Zhang, Y.: An empirical study of catastrophic forgetting in large language models during continual fine-tuning. IEEE Transactions on Audio, Speech and Language Processing (2025)

\bibitem{maclennan1996survey}
MacLennan, R.N., Peebles, J.W.E.: Survey of health problems and personality in Air Traffic Controllers. The International Journal of Aviation Psychology \textbf{6}(1), 43--55 (1996)

\bibitem{matzakos2023learning}
Matzakos, N., Doukakis, S., Moundridou, M.: Learning mathematics with large language models: A comparative study with computer algebra systems and other tools. International Journal of Emerging Technologies in Learning (iJET) \textbf{18}(20), 51--71 (2023)

\bibitem{meta_llama3_1}
Meta AI: Introducing Llama 3.1: Our most capable models to date (2024), \url{https://ai.meta.com/blog/meta-llama-3-1/}

\bibitem{national1980agenda}
{National Council of Teachers of Mathematics, Inc.}: An Agenda for Action: Recommendations for School Mathematics of the 1980s. National Council of Teachers of Mathematics (1980)

\bibitem{olive2010mathematical}
Olive, J., Makar, K., Hoyos, V., Kor, L.K., Kosheleva, O., Str{\"a}sser, R.: Mathematical knowledge and practices resulting from access to digital technologies. Mathematics Education and Technology-Rethinking the Terrain: The 17th ICMI Study, pp.~133--177. Springer (2010)

\bibitem{ouyang2022training}
Ouyang, L., Wu, J., Jiang, X., Almeida, D., Wainwright, C., Mishkin, P., Zhang, C., Agarwal, S., Slama, K., Ray, A., et al.: Training language models to follow instructions with human feedback. Advances in Neural Information Processing Systems \textbf{35}, 27730--27744 (2022)

\bibitem{polya2004solve}
Polya, G.: How to Solve It: A New Aspect of Mathematical Method. Princeton University Press (2004)

\bibitem{roscoe2007understanding}
Roscoe, R.D., Chi, M.T.H.: Understanding tutor learning: Knowledge-building and knowledge-telling in peer tutors' explanations and questions. Review of Educational Research \textbf{77}(4), 534--574 (2007)

\bibitem{scardamalia2006knowledge}
Scardamalia, M., Bereiter, C.: Knowledge building. The Cambridge (2006)

\bibitem{schoenfeld1981episodes}
Schoenfeld, A.H.: Episodes and executive decisions in mathematical problem solving. For the Learning of Mathematics \textbf{2}(2), 23--44 (1981)

\bibitem{schoenfeld2014mathematical}
Schoenfeld, A.H.: Mathematical Problem Solving. Elsevier (2014)

\bibitem{sweller1988load}
Sweller, J.: Load during problem solving. Cognitive Science \textbf{12}(2) (1988)

\bibitem{tambunan2019effectiveness}
Tambunan, H.: The Effectiveness of the Problem Solving Strategy and the Scientific Approach to Students' Mathematical Capabilities in High Order Thinking Skills. International Electronic Journal of Mathematics Education \textbf{14}(2), 293--302 (2019)

\bibitem{ukobizaba2021assessment}
Ukobizaba, F., Nizeyimana, G., Mukuka, A.: Assessment Strategies for Enhancing Students' Mathematical Problem-Solving Skills: A Review of Literature. Eurasia Journal of Mathematics, Science and Technology Education \textbf{17}(3) (2021)

\bibitem{wang2022super}
Wang, Y., Mishra, S., Alipoormolabashi, P., Kordi, Y., Mirzaei, A., Arunkumar, A., Ashok, A., Dhanasekaran, A.S., Naik, A., Stap, D., et al.: Super-NaturalInstructions: Generalization via declarative instructions on 1600+ NLP tasks. arXiv preprint arXiv:2204.07705 (2022)

\bibitem{wang2023instructuie}
Wang, X., Zhou, W., Zu, C., Xia, H., Chen, T., Zhang, Y., Zheng, R., Ye, J., Zhang, Q., Gui, T., et al.: InstructUIE: Multi-task instruction tuning for unified information extraction. arXiv preprint arXiv:2304.08085 (2023)

\bibitem{wang2024tutor}
Wang, R.E., Ribeiro, A.T., Robinson, C.D., Loeb, S., Demszky, D.: Tutor Copilot: A human-AI approach for scaling real-time expertise. arXiv preprint arXiv:2410.03017 (2024)

\bibitem{wardat2023chatgpt}
Wardat, Y., Tashtoush, M.A., AlAli, R., Jarrah, A.M.: ChatGPT: A revolutionary tool for teaching and learning mathematics. Eurasia Journal of Mathematics, Science and Technology Education \textbf{19}(7), em2286 (2023)

\bibitem{xu2023wizardlm}
Xu, C., Sun, Q., Zheng, K., Geng, X., Zhao, P., Feng, J., Tao, C., Jiang, D.: WizardLM: Empowering large language models to follow complex instructions. arXiv preprint arXiv:2304.12244 (2023)

\bibitem{yapatang2022development}
Yapatang, L., Polyiem, T.: Development of the Mathematical Problem-Solving Ability Using Applied Cooperative Learning and Polya's Problem-Solving Process for Grade 9 Students. Journal of Education and Learning \textbf{11}(3), 40--46 (2022)

\bibitem{zhou2024lima}
Zhou, C., Liu, P., Xu, P., Iyer, S., Sun, J., Mao, Y., Ma, X., Efrat, A., Yu, P., Yu, L., et al.: LIMA: Less is more for alignment. Advances in Neural Information Processing Systems \textbf{36} (2024)

\bibitem{zimmerman2002becoming}
Zimmerman, B.J.: Becoming a self-regulated learner: An overview. Theory into Practice \textbf{41}(2), 64--70 (2002)

\end{thebibliography}
